\documentstyle[epsf,prb,aps,preprint]{revtex}

\input epsf
\begin{document}
\draft

\title{Pairing, Stripes, Lattice Distortions and Superconductivity in 
Cuprate Oxides}
\author{A.~H.~Castro Neto}

\address{Department of Physics,
University of California,
Riverside, CA 92521}

\date{\today}
\maketitle

\begin{center}
\tt Proceedings of the MTSC 2000\\
Klosters, Switzerland \\
April 1, 2000.
\end{center}

\begin{abstract}
We propose a model for a spatially modulated collective state 
of superconducting cuprates
in which the electronic properties vary locally in space. 
In this model the regions
of higher hole density (called stripes) are described 
as Luttinger liquids and the regions of lower 
density (antiferromagnetic ladders) 
as an interacting bosonic gas of d$_{x^2-y^2}$ hole pairs.
We show that the transition to the superconducting state is 
topological and driven by decay processes
among these elementary excitations in the presence of vibrations. 
\end{abstract}

\newpage

\bigskip

\narrowtext

The experimental evidence for charge and spin inhomogeneities in superconducting
cuprates has been accumulating over the last few years \cite{experiments}. 
It is believed that these inhomogeneities are the result of strong competing 
electron-electron and electron-lattice interactions. The strong interactions
can be hinted from the fact that cuprates are directly related to Mott insulators 
and not to metals. There is an emerging
point of view that inhomogeneities may be at the heart of the superconducting 
problem and that an important part of the physics occurs in
real space. This should be contrasted with a Fermi liquid description which
is dominated by the Fermi surface. In this work we propose
a model for superconductivity which requires a dual description in terms of
real and momentum space.

We assume the existence of two main length scales in the problem.
At the short length scales the physics is dominated by the
strong Coulomb forces associated with the Mott insulator \cite{mott}. 
These forces lead to the quantum confinement of holes \cite{strings}.
An example of such confinement happens
when two holes are injected into an
antiferromagnet and 
there is a gain in magnetic and kinetic energy if they 
move together instead of separately. This leads to the formation of bound
states of two holes
which due to the same strong Coulomb forces and lattice
symmetry have d$_{x^2-y^2}$ symmetry \cite{strings}. 
On the other hand, in the presence of magnetic defects such as anti-phase domain
walls one-dimensional (1D) structures such as ``stripes'' can be stabilized due to the gain
of {\it single} hole kinetic energy \cite{sasha}.
Stripes are regions of larger carrier density which have properties
analogous to the ones found in low-dimensional interacting Fermi systems such as Luttinger liquids 
\cite{luttinger}.
In this case a long length scale associated with the inter-stripe
distance emerges leading to a phase with
long-range orientational order in the absence of translational order,
that is, a state similar to a classical liquid crystal \cite{lubensky}.
In fact, because the charge modulated state (CMS) is composed of electrons
we expect this modulated phase to be quantum in nature, that is, a 
quantum liquid crystal \cite{qualiqcrys}. 

Let us consider a static CMS induced along the $y$ direction
in the 2D system with periodicity $N a$ where $a$ is the lattice spacing.
We assume that the CMS is a superposition of a dilute homogeneous 
bosonic background of d$_{x^2-y^2}$
hole pairs in 2D and a set of 1D Luttinger liquids. 
This picture is supported by
density matrix renormalization group (DMRG) calculations 
of the $t-J$ model \cite{whitescalapino,elbio}. 
In fact, weak disorder \cite{nils},
spin-orbit effects and lattice distortions \cite{tranquada} might stabilize 
density modulations even further.
The CMS is such
that bosons are continuously created and destroyed
from the Luttinger liquids. In the absence of coupling between such
excitations the system should be insulating even in the presence of the
smallest amount of disorder: a Bose glass of pairs \cite{fisher} 
and a charge density wave state (CDW) \cite{luttinger} on the stripes
(we assume repulsive interactions only).  
Deformations of the amplitude of the density of
the CMS are always energetically costly. Therefore, phase fluctuations are the low energy excitations in such systems. 
Moreover, charge neutrality implies that a CMS has to be
coupled to the lattice. Therefore fluctuations of the CMS should
appear directly into the phonon spectrum. 
We believe that this has been observed experimentally in
essentially all the cuprates \cite{experphonons,egami}. 
The mechanism for superconductivity discussed in this work requires the coupling of
stripe fermions via the exchange of bosons. One can think of this mechanism as the exchange
of stripe ``pieces''. 
The simplest exchange mechanism is due to the decay of these bosons into
fermionic degrees of freedom at the stripes. 
We find, however, that the decay process
requires the production of lattice vibrations.
As a natural consequence, 
superfluid vortices are coupled to dislocations of the CMS. 
Because
the superfluid density is low, the interactions between 
topological defects ultimately determine the phase diagram \cite{jan}. 

We assume the existence of bosons with energy $E_{{\bf k}}$ and local repulsion $U$ described by
creation (annihilation) operators $P^{\dag}_{{\bf k}}$ ($P_{{\bf k}}$) and fermions
with energy dispersion $\epsilon_{{\bf k}}$ and spin $\sigma$ ($\uparrow$ or $\downarrow$)
described by creation (annihilation) operators $\psi^{\dag}_{{\bf k},\sigma}$ ($\psi_{{\bf k},\sigma}$).
When $N \gg 1$ the direct exchange of fermions between
stripes is suppressed and we can describe the fermions as isolated 1D
Luttinger liquids. 
In this limit it is convenient to label the fermion operators as $\psi_{n,\alpha,\sigma}(x)$
where $n$ is the stripe label, $\alpha=R,L$ refers to right and left moving fermions \cite{luttinger}.
On the one hand, bound states of two holes
are not eigenstates of the Luttinger liquid. On the other hand, collective sound
waves of the Luttinger liquid are not elementary excitations of the 
weakly doped Mott
insulator. Thus, the simplest process which couples these two types of excitations
is a decay process. The decay produces a ``shake'' of the
collective state emitting phonons. Conservation of linear and angular momentum requires the coupling
between fermions and bosons to have the form (we use units such that
$\hbar=k_B=1$ and $\beta=1/T$):
\begin{eqnarray}
H_c = \sum_{{\bf k},{\bf p},{\bf q}} V_{{\bf k}}({\bf q}) \lambda_{{\bf p}} 
P_{{\bf k}} \psi_{{\bf q},\uparrow} \psi_{-{\bf k}+{\bf p}-{\bf q},\downarrow} 
b^{\dag}_{{\bf p}} + h.c. \, . 
\label{coupling}
\end{eqnarray}
where $b_{{\bf p}}$ ($b^{\dag}_{{\bf p}}$) is the annihilation (creation) operators for
vibrations of the collective state (including the lattice). Here $\lambda_{p} \to 0$ when $p \to 0$ 
is the stripe-vibration coupling constant (the momentum dependence is a consequence of the
local coupling of the bosons with the stripes), 
$V_{{\bf k}}({\bf q})$ is the boson-stripe coupling which, by conservation of angular momentum,
has the d$_{x^2-y^2}$ symmetry. 
Notice that to destroy
a boson is equivalent to create two electrons in the antiferromagnetic 
ladder. 
In fact, (\ref{coupling}) is very similar to
fermion-boson models which have appeared in the literature of superconductivity
over the years \cite{rumer}. It is straightforward to see from (\ref{coupling})
that the condensation of the bosons with simultaneous condensation of fermions
into Cooper pairs with zero center of mass 
momentum requires superfluidity and lattice
distortions to occur at the same wave-vector. In particular, the largest
coupling occurs at the edge of the Brillouin zone with wave-vector 
${\bf K} = (0,\pm \pi/a)$ 
and thus perpendicular to the orientation of the charge
modulation (that is, perpendicular to the ``stripes''). 
In the case of the formation of domains with stripes running in
orthogonal directions we also expect
response at ${\bf K}^* = (\pm \pi/a,0)$.
If we forget about the phase fluctuations we would
find a superfluid/superconducting phase with simultaneous dimerization of
the CMS perpendicular to its orientation. 
This result would be in agreement with
the {\it simultaneous} 
appearance of the pseudogap in angle resolved photoemission experiments
\cite{arpes} (ARPES) and the phonon anomalies observed in neutron scattering
\cite{egami} exactly at ${\bf K}$. We stress that although vibrations are
not the mechanism of pairing (which is essentially produced
the electron-electron interactions), they are essential for condensation.

Because we are discussing a 2D system with low superfluid density the
problem of superconductivity is decided by phase fluctuations.
Consider the partition function of the problem which can be written
in terms of an Euclidean Lagrangian $L$.  We can separate the problem into fast
and slow moving fields by linearizing the fermion dispersion close to the 
stripe Fermi surface
and studying the bosons and lattice distortions close to ${\bf K}$.
We disregard all the fast oscillating terms in the Euclidean
action and rewrite the Lagrangian as
$L = L_0[P,b] + \sum_n L_S[\psi_n] + L_C$ where 
\begin{eqnarray}
L_0[P,b] &=& \frac{1}{2} \int d^2 r 
\left\{\overline{P}({\bf r}) \left(\partial_{\tau}+E_0-\frac{\nabla^2}{2 M_B}\right) P({\bf r}) 
+ \frac{U}{2} N^2 ({\bf r}) \right.
\nonumber
\\
&+& \left. \overline{b}({\bf r}) \left(\partial_{\tau}+\omega_0-\frac{\nabla^2}{2 M_V}\right) b({\bf r})\right\}
\end{eqnarray}
where $E_0$ ($\omega_0$) is the energy of the bosons (vibration) at ${\bf K}$, 
$M_B$ is the effective boson mass, 
$M_V$ is the mass associated with the vibrations and $N({\bf r})$ is
the local number of bosons, 
$L_S$ is the Lagrangian for interacting fermions in 1D \cite{luttinger} and
\begin{eqnarray}
L_C &=&  \frac{1}{2} 
\sum_{n} \int dx V(x,n) \left[P(x,n) \overline{b}(x,n)
+P(x,n) \overline{b}(x,n)\right]
\left[\psi_{n,R,\uparrow}(x) \psi_{n,L,\downarrow}(x) 
\right. 
\nonumber
\\
&+& \left. \left. \psi_{n,L,\downarrow}(x) \psi_{n,R,\uparrow}(x)\right]
+ c.c. \right\}
\end{eqnarray}
is the boson-stripe-vibration coupling. 
We now redefine fields as $P = \sqrt{\sigma_B} e^{-i \varphi_B}$ and
$b = \sqrt{\sigma_V} e^{+i \varphi_V}$
and use the bosonization technique \cite{luttinger} in order to get
$L = \sum_n L_{1D}[\theta_{n,\rho},\phi_{n,s}] + \sum_{\alpha=B,V} L_{\alpha}
[\sigma_{\alpha},\varphi_{\alpha}] + L_C[\sigma_{\alpha},\varphi_{\alpha},
\theta_{\rho},\phi_{s}]$
where $L_{1D}[\theta_{n,\rho},\phi_{n,s}]$ is the action for 1D bosons 
($\theta_{\rho}$ and $\phi_s$ are the charge and spin bosonic modes at
the stripes) \cite{luttinger} and
\begin{eqnarray}
L_{\alpha} = \int d^2r \left[i \sigma_{\alpha} \partial_{\tau} \varphi_{\alpha}+ E_{\alpha} \sigma_{\alpha} + \frac{U_{\alpha}}{2} \sigma^2_{\alpha}
 + \frac{\sigma_{\alpha}^{-1}}{8 M_{\alpha}}
(\nabla \sigma_{\alpha})^2 + \frac{\sigma_{\alpha}}{2 M_{\alpha}}
(\nabla \varphi_{\alpha})^2\right]
\end{eqnarray}
with $E_B = E_0$, $E_V = \omega_0$, $U_B = U$ and $U_V=0$. Finally,
\begin{eqnarray}
L_C = \sum_n \int \frac{dx}{2 \pi a} V(x,n N a)
\sqrt{\sigma_B \sigma_V} e^{-i (\varphi_B(x,n N a) + \varphi_V(x,n N a) - \sqrt{2 \pi} 
\theta_{n,\rho}(x))} \cos(\sqrt{2 \pi} \phi_{n,s}(x)) + c.c. \, .
\end{eqnarray}
We can gauge away the phase fields 
and define a new bosonic field $\theta_c$ such that
$\theta_{n,c}(x) = \theta_{n,\rho}(x) - (\varphi_B(x,n N a)+\varphi_V(x,n N a))/\sqrt{2 \pi}$
so that $L_C$ simplifies to
\begin{eqnarray}
L_C = \sum_n \int \frac{dx}{\pi a} V(x,n N a)
\sqrt{\sigma_B \sigma_V} \cos(\sqrt{2 \pi} \theta_{n,c}(x)) 
\cos(\sqrt{2 \pi} \phi_{n,s}(x)) \, .
\label{pairing}
\end{eqnarray}
Because of the gauge transformation a new term appears in the Lagrangian:
\begin{eqnarray}
L_I[\varphi] &=& \sum_n \int dx \left\{
\frac{K_{\rho}}{4 \pi v_{\rho}}
\left[ 
\left(\partial_{\tau}(\varphi_B(x,n)+\varphi_V(x,n))\right)^2 
%\right.
%\right.
%\nonumber
%\\
%&+& \left. \left. 
+ v_{\rho}^2
\left(\partial_{x}(\varphi_B(x,n)+\varphi_V(x,n))\right)^2 \right] \right.
\nonumber
\\
&+& \left. \frac{K_{\rho}}{\sqrt{2 \pi} v_{\rho}} 
\left[
\partial_{\tau}(\varphi_B(x,n)+\varphi_V(x,n)) 
\partial_{\tau}\theta_{n,c}(x) 
%\right.
%\right.
%\nonumber
%\\
%&+& \left. 
%\left.
+ v_{\rho}^2 
\partial_{x}(\varphi_B(x,n)+\varphi_V(x,n)) 
\partial_{x}\theta_{n,c}(x) \right] \right\} \, .
\label{lvarphi}
\end{eqnarray}
Here $K_{\rho}$ is the Luttinger parameter and 
$v_{\rho}$ the charge velocity.
We can study the phase fluctuations around the
saddle point where: $\sigma_{\alpha} 
= \sigma_{\alpha}^0$, $\varphi_{\alpha} = 0$ with $\alpha=B,V$. This just gives the mean
field solution which will be studied elsewhere \cite{next}. This mean field solution 
produces a characteristic temperature $T^*$ below which superfluidity and lattice distortions
start to be simultaneously generated at ${\bf K}$. 
This temperature, however, is not the actual transition temperature since
the phases of the order parameters fluctuate strongly. $T^*$ is a crossover temperature
which we associate with the so-called pseudogap phase of the cuprates \cite{loram}.
Around the saddle point we can
easily show that $L_C$ in (\ref{pairing}) is a relevant operator in
a renormalization group sense even in the presence
of strong repulsion. 
Thus, even if {\it isolated} stripes are deep inside of a CDW state
the coupling $V$ drives the system to a superconducting state. 
Amplitude fluctuations around the saddle point 
are therefore massive and produce a spin gap
in the system.
Integrating over amplitude fluctuations up to second order leads to
an Euclidean action of the form: 
\begin{eqnarray}
S &=& \frac{1}{2} \int_0^{\beta}d\tau  \int d^2 r  \left\{ 
\sum_{\alpha=B,V} \left[ 
i \overline{\sigma}_{\alpha} \partial_{\tau} \varphi_{\alpha}({\bf r},\tau) 
+ \frac{\kappa_{\alpha}}{4} \left(\partial_{\tau} \varphi_{\alpha}({\bf r},\tau)\right)^2
+ \frac{\sigma_{\alpha}}{M_{\alpha}} 
\left(\nabla \varphi_{\alpha}({\bf r},\tau)\right)^2 
\right]
\right.
\nonumber
\\
&+& \left. 
\frac{K_{\rho}}{2 \pi v_{\rho} N a} 
\left[\left(\sum_{\alpha=B,V} \partial_{\tau} \varphi_{\alpha}({\bf r},\tau)
\right)^2 + v_{\rho}^2 \left(\sum_{\alpha=B,V} \partial_x \varphi_{\alpha}({\bf r},\tau)
\right)^2 
\right] 
\right\}
\label{phaseaction}
\end{eqnarray}
where the factor of $1/N$ appears due to the coarse-graining of 
the fields in the 
direction perpendicular to the stripes. Here
$\overline{\sigma}_B$ is the total density of electrons
in the system, $\kappa_B$ is the
charge compressibility (the superfluid velocity is given by
$c_B = 4 \sigma_B/(\kappa_B M_B)$), $\overline{\sigma}_V =0$ and
$\kappa_V$ is the lattice compressibility (the sound velocity is
$c_V = 4 \sigma_V/(\kappa_V M_V)$).
The key point about (\ref{phaseaction}) is that the topological excitations of
the superfluid state (the vortices) are directly coupled 
to the topological excitations of the dimerized CMS (the 
dislocations loops) via the stripes.

Let us consider now the problem at $0<T<T^*$ so that the amplitude of
the order parameters are well developed but phase coherence has
not been established yet. At finite temperatures we can disregard the time derivatives
in (\ref{phaseaction}) and rewrite
\begin{eqnarray}
S = \frac{\beta}{2} \int d^2 r  \left\{ 
\sum_{\alpha=B,V}
\frac{\sigma_{\alpha}}{M_{\alpha}} 
\left(\nabla \varphi_{\alpha}({\bf r})\right)^2 
+  \frac{v_F}{2 \pi N a} 
\left(\sum_{\alpha=B,V} \partial_x \varphi_{\alpha}({\bf r})
\right)^2  
\right\} \, 
\label{phachigh}
\end{eqnarray}
which describes two 2D XY models coupled along the 
$x$ direction ($v_F$ is the stripe Fermi velocity). 
In what follows we will study the problem of phase coherence due to
the renormalization of one of the phase modes by the gaussian modes 
of the other.
This procedure works when the coupling is very weak, that is, 
$N \gg 1$. Because of the quadratic nature of the action it is a
simple exercise to show that the problem is described by a 2D XY model
with renormalized stiffness. In the superfluid case the stiffness is given
by
\begin{eqnarray}
\rho_B = (E_B+E_V) \left[1+\frac{E_V/E_B}{
\sqrt{1+E_C \left(E_B^{-1}+E_V^{-1}\right)}}\right]^{-1} 
\label{rhob}
\end{eqnarray}
where $E_{\alpha}=\sigma_{\alpha}/M_{\alpha}$ and $E_C=v_F/(2 \pi N a)$. 
An analogous result is obtained for the rigidity of the 
vibrations with the replacement of
of $E_B$ by $E_V$ and vice-versa.
Therefore, we conclude that the transition from the ordered to the
disordered phase of the superfluid or/and the dimerized 
CMS is due to the 
unbinding of the topological excitations: vortex-antivortex pairs
in the case of the superfluid and dislocation loops of the dimerized 
CMS. In
fact these topological excitations are coupled since a dislocation
of the dimerized CMS can potentially liberate vortex-antivortex pairs.
The transition temperature to the ordered phase can now be estimated
directly from (\ref{rhob}) as
\begin{eqnarray}
T_{c,\alpha} \approx \frac{\pi}{2} \rho_{\alpha} \, 
\label{tca}
\end{eqnarray}
in accord with the Uemura relation \cite{uemura}. 
An interesting consequence of our calculations is that if the phonon at 
${\bf K}$ is optical (dispersionless) then $E_V \to 0$ 
and $\rho_V \to 0$. In this case the
transition to the CMS is driven to zero
temperature! Thus, while the system becomes a superfluid-superconductor
at $T_{c,B}$, at any finite temperature dislocations loops of the 
dimerized CMS will render $T_{c,V}=0$.

In summary, we have proposed a 2D model for superconducting order in
cuprates which involve a spatially modulated collective state which
contains d$_{x^2-y^2}$ bosons and stripe fermions. We show that the
coupling between these particles leads to superconductivity and 
simultaneous condensation
of bosons and phonons at finite wave-vector. 
Although condensation starts at high temperature $T^*$, phase coherence
is only attained at low temperatures due to proliferation of vortex-antivortex
pairs of the superfluid and dislocation loops of a dimerized CMS.

I would like to acknowledge fruitful conversations with A.~Balatsky, 
S.~Billinge, A.~R.~Bishop, A.~Chernyshev, T.~Egami, L.~Pryadko, 
R.~McQueeney, D.~MacLaughlin and J.~Zaanen. 
I acknowledge partial support from a Los Alamos CULAR grant under the 
auspices of the U.~S. Department of Energy.

\end{document}